\documentclass[aps,prd, nofootinbib]{revtex4}

\usepackage{amsmath, amsfonts, amsthm, amssymb, graphicx}
\usepackage{subfig}
\usepackage{epstopdf}

\def\be{\begin{equation}}
\def\ee{\end{equation}}
\def\ba{\begin{eqnarray}}
\def\ea{\end{eqnarray}}

\def\bdm{\begin{displaymath}}
\def\edm{\end{displaymath}}

\def\ga{~\mbox{\raisebox{-.6ex}{$\stackrel{>}{\sim}$}}~}
\def\bq{\begin{quote}}
\def\eq{\end{quote}}

\def\ga{~\mbox{\raisebox{-.6ex}{$\stackrel{>}{\sim}$}}~}

\def\del{\partial}
\def\ltap{\ \raise.3ex\hbox{$<$\kern-.75em\lower1ex\hbox{$\sim$}}\ }
\def\gtap{\ \raise.3ex\hbox{$>$\kern-.75em\lower1ex\hbox{$\sim$}}\ }
\def\gl{\ \raise.5ex\hbox{$>$}\kern-.8em\lower.5ex\hbox{$<$}\ }
\def\roughly#1{\raise.3ex\hbox{$#1$\kern-.75em\lower1ex\hbox{$\sim$}}}

 at 10truept

\newcommand{\beq}{\begin{equation}}
\newcommand{\eeq}{\end{equation}}
\newcommand{\bea}{\begin{eqnarray}}
\newcommand{\eea}{\end{eqnarray}}
\newcommand{\beqa}{\begin{eqnarray}}
\newcommand{\eeqa}{\end{eqnarray}}
\newcommand{\nn}{\nonumber\\}

\def \pd {\partial}

\begin{document}

\title{Bi-galileon theory I: motivation and formulation}

\author{Antonio Padilla} 
\email[]{antonio.padilla@nottingham.ac.uk}
\author{Paul M. Saffin} 
\email[]{paul.saffin@nottingham.ac.uk}
\author{Shuang-Yong Zhou} 
\email[]{ppxsyz@nottingham.ac.uk}

\affiliation{School of Physics and Astronomy, 
University of Nottingham, Nottingham NG7 2RD, UK} 

\date{\today}

\begin{abstract}
We introduce bi-galileon theory, the generalisation of the single galileon model introduced by Nicolis {\it et al}. The theory contains two coupled scalar fields  and is described by a Lagrangian that is invariant under Galilean shifts in those fields. This paper is the first of two, and focuses on the motivation and formulation of the theory. We show that the boundary effective theory of the cascading cosmology model corresponds to a bi-galileon theory in the decoupling limit, and argue that this is to be expected for co-dimension 2 braneworld models exhibiting infra-red modification of gravity.   We then generalise this, by constructing the most general bi-galileon Lagrangian. By coupling one of the galileons to the energy-momentum tensor, we pitch this as a modified gravity theory in which the modifications to General Relativity are encoded in the dynamics of the two galileons. We initiate a study of phenomenology by looking at maximally symmetric vacua and their stability, developing elegant geometric techniques that trivially explain why some  of the vacua have to be unstable in certain cases (eg DGP). A detailed study of phenomenology appears in our companion paper. 

\end{abstract}


\maketitle

\section{Introduction}
When Urbain Le Verrier first noticed that the slow precession of Mercury's orbit could not be explained using perturbative Newtonian gravity and the known planets, he postulated the existence of a hitherto unseen {\it dark} planet, which he called {\it Vulcan}~\cite{vulcan}. Of course,  thanks to Einstein, we now know that what was really required was a new theory of gravity,  specifically, the General Theory of Relativity~\cite{GR}. 150 years later, flat rotations curves of galaxies~\cite{rubin} and accelerated cosmic expansion~\cite{acceleration} have led most of us to follow Le Verrier's  approach and to postulate the existence of dark matter~\cite{dm} and dark energy~\cite{de} respectively.  For dark matter, bullet cluster observations suggest that Le Verrier's approach is probably the right one~\cite{bullet}. The Le Verrier approach to  dark energy is less compelling. Whereas extensions of the Standard Model  offer some very good dark matter candidates, they have been nothing short of hopeless in offering a satisfactory candidate for dark energy. It is  certainly worth taking a lesson from history and exploring the possibility that a new theory of gravity may be required to explain the observed cosmic acceleration.

In order to play a role in the accelerated expansion, we need gravity to deviate from General Relativity (GR) at large scales,  and in particular at the scale of the cosmic horizon $H_0^{-1} \sim 10^{26} \mathrm{m}$.  Such deviations should not be felt all the way down to arbitrarily short distances because General Relativity is very successful in describing classical gravity up to the scale of the Solar System, at about $10^{15} \mathrm{m}$.  Modifications of General Relativity are most easily understood at the level of linearised theory. In GR, the gravitational force is mediated by a massless spin 2 particle carrying two polarisation degrees of freedom. In modified gravity we expect additional degrees of freedom to be present.  For example,  in Pauli-Fierz massive gravity the graviton carries 5, rather than 2, polarisation degrees of freedom~\cite{pf}. Clearly  if they are to play a role in accelerated expansion without spoiling the phenomenological successes of GR at shorter distances, the additional degrees of freedom should be suppressed up to Solar System scales, but not on cosmological scales. This should occur without introducing any new pathological modes. In  massive gravity,  the longitudinal graviton mode is problematic  since it gives rise to the so called vDVZ discontinuity in the graviton propagator, even at scales where the graviton mass is negligible~\cite{vdvz}. However, this mode can be "screened"   on Solar System scales thanks to non-linear interactions and the Vainshtein mechanism~\cite{vainshtein}, but only at the expense of introducing a new pathology in the form of the Boulware-Deser ghost \cite{bdghost, sc-ghosts}.

Extra dimensions and the braneworld paradigm~\cite{bw} have opened up some intriguing possibilities for modifying gravity at large distances (see, for example, \cite{dgp, kogan, ruth, me}). The DGP model~\cite{dgp} is probably the most celebrated example of this. The model admits two distinct sectors: the normal branch and the self-accelerating branch. The latter generated plenty of interest since it gave rise to cosmic acceleration without the need for dark energy~\cite{sa}. However, fluctuations about the self-accelerating vacuum were found to suffer from ghost-like instabilities, ruling out this sector of the theory~\cite{saghosts, luty}.  Although the normal branch is less interesting phenomenologically, it is fundamentally more healthy and is the closest thing we have to a consistent non-linear completion of massive gravity.   Here the graviton is a resonance of finite width $H_0$, as opposed to a massive field. At short distances, $r \ll H_0^{-1}$ the brane dynamics does not feel the width of the resonance and the theory resembles $4D$ GR. At large distances, $r \ga H_0^{-1}$, however, the theory becomes five dimensional as the resonance effectively decays into continuum Kaluza-Klein modes.  The Vainshtein mechanism works well on the normal branch of DGP, screening the longitudinal graviton without introducing any new pathological modes, in contrast to massive gravity~\cite{sc-ghosts, dgp-vainsh}. 

Much of the interesting dynamics is contained in the properties of the longitudinal scalar, $\pi$. One can identify this mode with the brane bending mode. As this mode only excites the extrinsic curvature of the brane, and not the brane or bulk geometry, it becomes strongly coupled well below the $4D$ Planck scale~\cite{luty}. It is precisely this strong coupling that allows the Vainshtein mechanism to take effect as non-linearities become important at larger than expected distances, helping to shut down the scalar on Solar System scales~\cite{dvalithm, sc-ghosts}. We can isolate the dynamics of the scalar by taking the so-called decoupling limit around flat space, in which the $4D$ and $5D$ Planck scales go to infinity, but the strong coupling scale is held fixed. In this limit  the tensor mode, $h_{\mu\nu}$, decouples from the scalar, $\pi$, to all orders~\cite{luty,nicolis}. The $\pi$-Lagrangian now contains a plethora of information about both branches of solution, which can be studied much more easily than in the full theory.

Now an important feature of the $\pi$-Lagrangian is that it  exhibits {\it galileon} invariance~\cite{gal}. That is, the action is invariant under $\pi \to \pi +b_\mu x^\mu+c$, where $b_\mu$ and $c$ are constant. This essentially follows from the fact that the brane bending contribution to the extrinsic curvature is given by $K_{\mu\nu} \simeq \del_\mu\del_\nu  \pi$.  Equivalently, Poincare transformations in the bulk  should not affect the physics of the brane bending mode. If the brane position is given by $y=\pi(x)$,  then after an infinitesimal Poincare transformation, $x^a \to x^a+\epsilon^a{}_b x^b+\upsilon^a$ the brane position is given by $y \simeq \pi(x)+\epsilon^y{}_\mu x^\mu+\upsilon^y$. Since the physics of the brane bending mode should remain unaffected it is clear that it should exhibit galileon invariance.

Motivated by the DGP model, Nicolis {\it et al}~\cite{gal} developed the idea of galileon invariance and wrote down the most general Lagrangian for a single galileon degree of freedom (see \cite{david} for some early work along similar lines). Remarkably the theory contains only five free parameters, enabling the authors, and others \cite{clare}, to complete a thorough analysis.  Although the Galilean symmetry is broken when we include coupling to gravity~\cite{covgal}, covariant completions of the general galileon action can be identified with the position modulus of a probe DBI brane~\cite{dbigal}.  In any event, it seems clear that any co-dimension one brane model with large distance  deviations from GR, will indeed admit a galileon description around flat space, in some appropriate limit. This follows from  the discussion of the brane bending mode in the previous paragraph, and from~\cite{dvalithm}, where it is argued that any viable infra-red modification of gravity leads to strong coupling, and is therefore expected to admit a non-trivial decoupling limit. 

Recently there has been plenty of interest in branes of higher co-dimension and the role they might play in modifying gravity at cosmological scales~\cite{casc1, casc2, casccos,  koyama, cline, SLED, nk, ccpap, cczzz, papa}. An example of this  is the cascading DGP model~\cite{casc1}, in which one has DGP branes within DGP branes~\cite{casc1, casc2, casccos}. In the $6D$ version one has a co-dimension 2 DGP 3-brane intersecting a co-dimension one DGP 4-brane~\cite{casc1}.  Higher co-dimension brane models seem to have extra features, such as degravitation~\cite{degrav, casc2}, or perhaps even self-tuning~\cite{casc1,koyama, cline, SLED}, in which the geometry of the 3-brane is insensitive to a large 3-brane vacuum energy.   Now to specify the position of a co-dimension-$N$ brane we need to specify the values of  $N$ ``bulk" components: $y_1=\pi_1(x), y_2=\pi_2(x), \ldots,  y_N=\pi_N(x)$.  For a viable model giving rise to  large distance deviations from GR, there should be strong coupling and therefore an appropriate limit in which the brane bending modes decouple from the tensor mode around flat space. Then the physics of each brane bending mode should be unaffected by Poincare transformations in the bulk, and we conclude that we are left with an $N$-galileon theory.  

In this paper we will focus on the case of two coupled galileons, which we will call {\it bi-galileon} theory. Of course, this has particular relevance to co-dimension 2 brane models. Indeed, in section \ref{sec:cascosmo}, we will show explicitly that the $5D$ cascading cosmology model is one example of a bi-galileon theory.  This model is derived from the original $6D$ cascading gravity model \cite{casc1, casc2} by computing the  boundary effective field theory in $5$ dimensions, and taking a suitable decoupling limit.  Although a  detailed study of phenomenology will appear elsewhere~\cite{otherpaper}, this paper is devoted to the motivation and formulation of the most general bi-galileon theory.  In section \ref{sec:modgrav} we will pitch our  theory as a modification of gravity by coupling one of the galileon fields to the energy-momentum tensor. We will not couple the graviton directly to the scalars, which amounts to ignoring their backreaction on to the geometry, as was done for the case of the single galileon~\cite{gal}. In section \ref{sec:gentheory} we will construct the general bi-galileon theory. We will also state the extension to any number of coupled galileon fields. We initiate a study of phenomenology in section \ref{sec:maxsym} by looking at maximally symmetric vacua and their stability. Section \ref{sec:discuss} contains a half-time analysis.
\section{Example: cascading cosmology} \label{sec:cascosmo}

There have been a number of attempts to extend the DGP model to higher  dimensions. Although the earliest of these ran into problems with ghosts and instabilities~\cite{higherdgp}, the recently developed cascading models~\cite{casc1, casc2, casccos} have been more successful. These models exhibit degravitation and/or self-tuning, and seem to be free from ghosts, at least if the 3-brane tension is large enough. In this section we will compute the $4D$ boundary effective theory for the cascading cosmology model~\cite{casccos}. By taking the appropriate limit we will decouple the scalar sector and show that it corresponds to  a bi-galileon theory. 

The cascading cosmology model~\cite{casccos} is closely related to the $6D$ cascading DGP model~\cite{casc1}. The latter contains a DGP 3-brane sitting on a DGP 4-brane in 6 dimensions. The action is given by
\be
S_{cascading}=\frac{M_6^4}{2} \int_{bulk} d^6 x \sqrt{-g_6}~R_6+\frac{M_5^3}{2} \int_{4-brane} d^5 x \sqrt{-g_5}~R_5 
+\frac{M_4^2}{2} \int_{3-brane} d^4 x \sqrt{-g_4}~R_4\,.
\ee
If we have a hierarchy of scales $M_6 < M_5 < M_4$, then the Newtonian potential cascades from $1/r$ to $1/r^2$ to $1/r^3$ as we move further away from the source. The behaviour crosses over from $4D$ to $5D$ at a scale $m_5^{-1}$ and from $5D$ to $6D$ at a scale $m_6^{-1}$, where
\be
m_5=\frac{2M_5^3}{M_4^2}\,, \qquad m_6=\frac{2M_6^4}{M_5^3}\,.
\ee
In the original model~\cite{casc1} both crossover scales are taken to be of order the cosmic horizon, so that $m_5 \sim m_6 \sim H_0 \sim 10^{-33}$ eV. Taking $M_4 \sim M_{pl} \sim 10^{18}$ GeV, it follows that $M_5 \sim 10$ MeV and $M_6 \sim$ meV. Note that bulk quantum gravity corrections are expected to kick in at $M_6^{-1} \sim 0.1$ mm.

Now, one can integrate out the bulk, just as in \cite{luty}, to obtain the effective $5D$ theory.   It turns out that the longitudinal mode of the $5D$ graviton becomes strongly coupled at a scale
\be
\Lambda_6=(m_6^4 M_5^3)^{1/7} \sim 10^{-16} ~\textrm{eV}\,,
\ee
or, equivalently, at distances of order $\Lambda_6^{-1} \sim 10^6$ km.  Now at distances much shorter than the cosmic horizon, but much larger than the millimeter scale,  we can happily take $M_5, ~M_6 \to \infty$, whilst holding $\Lambda_6$ fixed. This corresponds to the decoupling limit in the effective $5D$
theory~\cite{casc1, casccos}. The cascading cosmology model~\cite{casccos} represents a non-linear completion of this decoupled effective theory\footnote{This non-linear completion is certainly not unique but one can reasonably expect it capture the salient features of the original $6D$ theory.}. The action is given by 
\begin{align}
S=&~ M_5^3\int_{\cal M} d^5 x \sqrt{-g}~e^{-3\pi/2} R_5-\alpha (\del \pi)^2 \square_5 \pi
\\
& -2M_5^3 \int_{\del \cal M} d^4 x \sqrt{-q} ~\left[ e^{-3\pi/2} K+\frac{\alpha}{2}\left( (\del_\mu \pi)^2 {\cal L}_n \pi +\frac{1}{3} ( {\cal L}_n \pi)^3 \right)\right]
+\frac{M_4^2}{2} \int_{\del \cal M} d^4 x \sqrt{-q}~R_4 \,, \label{cascact}
\end{align}
where the  full $5D$ spacetime is made up of two copies of $\cal M$, identified across a common boundary, $\del \cal M$, given by the 3-brane. Here $g_{ab}dx^a dx^b$ and $q_{\mu\nu}dx^\mu dx^\nu$ are the bulk and brane metric respectively,  with corresponding Ricci scalars  $R_5$ and $R_4$. The extrinsic curvature of the brane is given by $K_{\mu\nu}=\frac{1}{2}{\cal L}_n q_{\mu\nu}$ where   ${\cal L}_n$ is the Lie derivative with respect to the inward pointing normal, $n^a$, to $\del \cal M$.  The bulk scalar $\pi(x)$ is the $5D$ longitudinal mode and the parameter $\alpha=\frac{27}{4 m_6^2}$.

If we choose coordinates $x^a=(y,x^\mu)$, so that the brane is at $y=0$ and the bulk extends into $y>0$, then it is convenient to write the bulk metric in ADM coordinates
\be
ds^2=g_{ab}dx^a dx^b=N^2 dy^2+q_{\mu\nu}(dx^\mu +N^\mu dy)(dx^\nu +N^\nu dy)\,,
\ee
where $N$ is the lapse function and $N^\mu$ is the shift vector. Then 
\be
K_{\mu\nu}=\frac{1}{2N} \left(\del_y q_{\mu\nu}-2D_{(\mu} N_{\nu)} \right)\,, \qquad {\cal L}_n \pi=\frac{1}{N} \left(\del_y -N^\mu D_\mu \right)\pi\,,
\ee
where $D_\mu$ is the covariant derivative with respect to the brane metric $q_{\mu\nu}$.

To calculate the boundary effective action, we essentially follow steps completely analogous to the ones taken in \cite{luty}. In order to integrate out the bulk fields we must choose a gauge. We therefore set 
\be
g_{ab}=\eta_{ab}+h_{ab}\,, \qquad q_{\mu\nu}=\eta_{\mu\nu}+h_{\mu\nu}\,,
\ee
and add a bulk gauge fixing term
\be
S_{bulk, gf}=-\frac{M_5^3}{2}\int_{\cal M} d^5 x ~ (F_a+\frac{3}{2} \del_a \pi) (F^a+\frac{3}{2} \del^a \pi)\,,
\ee
where 
\be
F^a=\del_b h^{ab}-\frac{1}{2} \del^a h\,, \qquad h=\eta^{ab} h_{ab}\,.
\ee
Classically this imposes the gauge $F_a=-\frac{3}{2} \del_a \pi$. To quadratic order we have
\begin{align}
\delta_2 S=&~M_5^3 \int d^4 x dy~\frac{1}{4}\left(h^{ab}-\frac{1}{2} h\eta^{ab}\right)\square_5 h_{ab}+\frac{3}{4}\pi \square_5 h+\frac{9}{8}\pi \square_5 \pi \nonumber
\\
&+M_5^3  \int_{y=0} d^4 x ~\frac{1}{4}\left(h^{\mu\nu}- \frac{1}{2}h_4\eta^{\mu\nu}\right)\del_y h_{\mu\nu}+\frac{1}{2}N^\mu \del_{y} N_\mu +\frac{3}{4} \pi\del_y h_4-\frac{1}{8}(h_{yy} \del_y h_4+h_4 \del_y h_{yy})  \nonumber
\\
& ~~~~~~~~~~~~~~~~~~~~+\frac{1}{8} h_{yy} \del_yh_{yy}+\frac{3}{4}\pi \del_y h_{yy} +\frac{9}{8} \pi \del_y \pi+N_\mu\left( f^\mu-\frac{1}{2} \del^\mu h_{yy}+\frac{3}{2} \del^\mu \pi\right)
 \nonumber
\\
&+\frac{M_4^2}{2} \int_{y=0} d^4 x ~\frac{1}{4}\left(h^{\mu\nu}-\frac{1}{2} h_4\eta^{\mu\nu}\right)\square_4 h_{\mu\nu}+\frac{1}{2} f_\mu f^\mu\,,
\end{align}
where 
\be
f^\mu=\del_\nu h^{\mu\nu}-\frac{1}{2} \del^\mu h_4\,, \qquad
h_4=\eta^{\mu\nu}h_{\mu\nu}\,,
\ee
and we have re-defined $N_\mu=h_{\mu y}$. The bulk equations of motion
\be
\square_5 h_{ab}=0\,, \qquad \square_5 \pi=0
\ee
have solutions
\be
h_{ab}=e^{-y \Delta} h_{ab}\vert_{y=0}\,, \qquad \pi=e^{-y \Delta}\pi\vert_{y=0}\,,
\ee
where $\Delta=\sqrt{-\square_4}$. Now we integrate out the bulk and fix the residual gauge freedom left on the brane by adding a brane gauge fixing term,
\be
S_{brane, gf}=-\frac{M_4^3}{4}\int_{\cal M} d^4 x ~ (f_\mu+m_5 N_\mu) (f^\mu+m_5 N^\mu) \,.
\ee
This gauge fixing eliminates the mixing between  $N_\mu$ and $f_\mu$ and we are left with
\begin{align}
\delta_2 S=\frac{M_4^2}{2} \int_{y=0} d^4 x ~ & \frac{1}{4}\left(h^{\mu\nu}-\frac{1}{2} h_4\eta^{\mu\nu}\right)(\square_4-m_5 \Delta) h_{\mu\nu}
-m_5\left[ \frac{1}{2}N^\mu (\Delta+m_5) N_\mu 
\right.\nonumber \\ 
  & ~~~~~~~\left. +\frac{1}{2} N_{\mu}\del^\mu 
\left(h_{yy}-3 \pi\right)-\frac{1}{4}(h_{yy}-3\pi) \Delta h+\frac{1}{8} (h_{yy} +3\pi)\Delta (h_{yy}+3\pi)\right]\,. 
\end{align}
 We can diagonalise this action by making the following field re-definitions
\be
h_{\mu\nu}=\tilde h_{\mu\nu}+(\pi+\phi) \eta_{\mu\nu}\,, \qquad N_\mu=\tilde N_\mu+\frac{1}{m_5}\del_\mu (\pi+\phi)\,, \qquad h_{yy}=3\pi-\frac{2}{m_5} (\Delta+m_5)(\pi+\phi)\,. \label{redef}
\ee
This gives
\begin{align}
\delta_2 S= \frac{M_4^2}{2} \int_{y=0} d^4 x ~&\frac{1}{4}\left(\tilde h^{\mu\nu}-\frac{1}{2} \tilde h_4\eta^{\mu\nu}\right)(\square_4-m_5 \Delta) \tilde h_{\mu\nu}-\frac{1}{2} m_5 \tilde N^\mu (\Delta+m_5) \tilde  N_\mu \nn
&+\frac{3}{2}\phi (\square_4-m_5 \Delta)\phi-\frac{3}{2}\pi(\square_4+2m_5\Delta)\pi\,.
\end{align}
Note that the $\pi$ field is a ghost. This can be avoided by increasing the vacuum energy on the brane~\cite{casc1}. In addition, to quadratic order, the matter coupling is given by
\be \label{zmatteraction}
\delta_2 S_{matter}=\frac{1}{2} \int_{y=0} d^4 x h_{\mu\nu} T^{\mu\nu}=\frac{1}{2} \int_{y=0} d^4 x~ \tilde h_{\mu\nu} T^{\mu\nu}+(\pi+\phi)T\,.
\ee
We now consider higher order terms. Schematically it is clear from Eq.~\ref{cascact} that these will take the form
\bea
&&\int d^4 x ~M_5^3 \Delta (\pi)^{a_1}(h_{\mu\nu})^{b_1} (N_\mu)^{c_1} (h_{yy})^{d_1}\,,\label{higher}\\
&&\int d^4 x ~M_5^3\alpha  \Delta^3 (\pi)^{a_2} (h_{\mu\nu})^{b_2} (N_\mu)^{c_2} (h_{yy})^{d_2}\,, \label{a2}\\
&&\int d^4 x ~M_4^2 \Delta^2 (h_{\mu\nu})^{b_3}\,,
\eea
where $a_i, \ldots, d_i$ are non-negative integers satisfying $a_i+b_i+c_i+d_i \geq 3$.  Note that we must have $a_2 \geq 1$ since terms of the form Eq. \ref{a2} stem from the boundary $\pi$ term given in Eq. \ref{cascact}. After canonically normalising the fields, it is easy enough to check that the largest contribution comes from a term like Eq.~\ref{higher}, with $a_1=b_1=0, ~c_1+d_1=3$. This term is exactly like the cubic term computed in \cite{luty}, and comes from the following term in the action (\ref{cascact}) 
\be
M_5^3\int_{\cal M} d^5 x \sqrt{-g}~e^{-3\pi/2} R_5-2M_5^3 \int_{\del \cal M} d^4 x \sqrt{-q} e^{-3\pi/2} K\,.
\ee
Since we want $a_1=0$, we can switch off $e^{-3\pi/2}$, and write this term in ADM form,
\be
M_5^3 \int d^4 x dy \sqrt{-q}N( R_4 +K^2-K_{\mu\nu}K^{\mu\nu})\,.
\ee
Of course,  we also want $b_1=0$, so we need to switch off $h_{\mu\nu}$. This means $q_{\mu\nu}=\eta_{\mu\nu}$, and so $R_4=0$ and $K_{\mu\nu}=-\frac{1}{N} \del_{(\mu} N_{\nu)}$. Given that $N \approx 1+\frac{1}{2} h_{yy}$, the cubic order interaction term is given by
\be
-\frac{M_5^3}{2}\int d^4 x dy h_{yy}\left[ (\del_\mu N^\mu)^2-\del_{(\mu} N_{\nu)}\del^{\mu} N^{\nu}\right]\,. \label{cubic}
\ee
In the limit where $m_5 \to 0$, the strongly interacting mode can be parametrized {\it in the bulk} as
\be
h_{\mu\nu} \to 0\,, \qquad N_{\mu} \sim \del_\mu \Psi\,, \qquad h_{yy} \sim 2 \del_y \Psi\,, \qquad \pi\to 0\,,
\ee 
where $\square_5 \Psi=0$. This satisfies the bulk equations of motion, as well as the gauge choice $F_a=-\frac{3}{2}\del_a \pi$. Indeed, the strongly interacting mode  corresponds to a trivial bulk geometry, as  expected~\cite{luty}. Furthermore, we see from Eq.~\ref{redef} that the boundary values for the fields go like $N_\mu \sim \frac{1}{m_5}\del_\mu (\pi+\phi)$,  and $h_{yy} \sim -\frac{2}{m_5} \Delta(\pi+\phi) $ so we take $\Psi=e^{-\Delta y}\left(\frac{\pi+\phi}{m_5}\right)$. Eq.~\ref{cubic} can now be written as
\begin{align}
-M_5^3 \int d^4 x dy~ \del_y \Psi \left[ (\square_4 \Psi)^2 - (\del_\mu \del_\nu \Psi)^2\right] &=\frac{M_5^3}{2} \int d^4 x dy~ \del_y \left[ \del_\mu \Psi \del^\mu \Psi \square_4 \Psi\right] \nn
&=-\frac{M_5^3}{2m_5^3} \int d^4 x~ \del_\mu (\pi+\phi) \del^\mu(\pi+\phi) \square_4 (\pi+\phi)\,.
\end{align}
Now if we canonically normalise, so that $\pi=\sqrt{\frac{2}{3}}\frac{\hat \pi}{M_4}$ and  $\phi=\sqrt{\frac{2}{3}}\frac{\hat \phi}{M_4}$, then we see that the cubic interaction term goes like
\be
-\left(\frac{1}{3\sqrt{6}}\right)\frac{1}{ \Lambda_5^3 } \int d^4 x~ \del_\mu (\hat \pi+\hat \phi) \del^\mu(\hat \pi+\hat \phi) \square_4 (\hat \pi+\hat \phi)\,, \label{interaction}
\ee
where 
\be
\Lambda_5 = (m_5^2M_4)^{1/3} \sim 10^{-13} \textrm{eV}\,.
\ee
This represents the scale at which the $4D$ scalars becomes strongly coupled, and corresponds to a distance $\Lambda_5^{-1} \sim 1000$km. We can now take the formal limit\footnote{In taking this limit we inevitably project out the pure DGP scenario, since that would require us to take $M_6 \equiv 0, \Lambda_6 \equiv 0$, which is, of course incompatible with Eq. \ref{declimit}.} 
\be \label{declimit}
M_4, M_5, M_6, T_{\mu\nu} \to \infty\,, \qquad \Lambda_5= \textrm{const}\,, ~\Lambda_6=\textrm{const}, ~\frac{T_{\mu\nu}}{M_4}=\textrm{const}\,.
\ee
Note that this requires that  $m_5, m_6 \to 0$.  It is easy to see that the leading order matter coupling remains. In contrast, the large interaction term Eq.~\ref{interaction} is the only interaction term to survive this limit. As a result,  the tensor $\tilde h_{\mu\nu}$ and vector $\tilde N_{\mu}$ decouple from one another and from the scalars $\pi$ and $\phi$ to all orders. In terms of the canonically normalised fields, the action is given by
\begin{align} \label{deccascact}
S_{dec}=\int d^4x~&\frac{1}{2} \left(\hat h^{\mu\nu}-\frac{1}{2}\hat h_4 \eta^{\mu\nu}\right)\square_4 \hat h_{\mu\nu}-\frac{1}{2} \hat N^\mu \Delta \hat N_\mu+\frac{1}{2} \hat \phi \square_4\hat \phi-\frac{1}{2} \hat \pi\square_4\hat \pi\nn
&-\left(\frac{1}{3\sqrt{6}}\right) \frac{1}{\Lambda_5^3}  \del_\mu (\hat\pi+\hat\phi) \del^\mu(\hat\pi+\hat\phi) \square_4 (\hat\pi+\hat\phi)
+\hat h_{\mu\nu} \frac{T^{\mu\nu}}{M_4}+\frac{1}{\sqrt{6}} (\hat\pi+\hat\phi)\frac{T}{M_4}\,,
\end{align}
where $\hat h_{\mu\nu}=\frac{M_4}{2} \tilde h_{\mu\nu}$ and  $\hat N^\mu=M_5^{\frac{3}{2}} \tilde N^{\mu}$. Focussing on the scalar sector, the relevant piece of the action is given by
\be \label{casdecaction}
S_{\pi, \phi}=\int  d^4 x~ \frac{1}{2} \hat \phi \square_4\hat \phi-\frac{1}{2} \hat \pi\square_4\hat \pi-\left(\frac{1}{3\sqrt{6}}\right) \frac{1}{\Lambda_5^3}  \del_\mu (\hat\pi+\hat\phi) \del^\mu(\hat\pi+\hat\phi) \square_4 (\hat\pi+\hat\phi)+\frac{1}{\sqrt{6}} (\hat\pi+\hat\phi)\frac{T}{M_4}\,.
\ee
The resulting equations of motion are explicitly Galilean invariant,
\bea
\square_4 \hat \phi-\left(\frac{2}{3\sqrt{6}}\right) \frac{1}{\Lambda_5^3}  \left[(\del_\mu\del_\nu (\hat\pi+\hat\phi))^2-(\square_4 (\hat\pi+\hat\phi))^2\right]+\frac{1}{\sqrt{6}}\frac{T}{M_4}=0\,,&& \\
-\square_4 \hat \pi-\left(\frac{2}{3\sqrt{6}}\right) \frac{1}{\Lambda_5^3}  \left[(\del_\mu\del_\nu (\hat\pi+\hat\phi))^2-(\square_4 (\hat\pi+\hat\phi))^2\right]+\frac{1}{\sqrt{6}}\frac{T}{M_4}=0\,.&& 
\eea
That is, they are invariant under
\be
 \hat \pi \to \hat \pi+ b_\mu x^\mu+ c\,,\qquad \hat \phi \to \hat \phi+\tilde b_\mu x^\mu+\tilde c\,,
\ee
for constant $b_\mu, \tilde b_\mu, c$ and  $\tilde c$~\cite{gal}. Thus we confirm what we expected. Cascading cosmology admits a bi-galileon description in the full decoupling limit. The limit corresponds to taking the Planck masses to infinity but holding the $4D$ and $5D$ strong coupling scales fixed.  The latter correspond to $1000$ km and $1000000$ km respectively, whilst the Planck lengths in the original model are all submillimeter.  In taking the decoupling limit we have sent all the next largest interactions to zero, so our description is only valid at scales where these additional interactions may be ignored, up to the horizon scale. These interactions may have an important role to play in understanding various aspect of the full theory (eg.  Vainshtein effects), but since our goal was merely to demonstrate that a bi-galileon description exists in principle at certain scales, we do not concern ourselves with that here.
%


The $\pi,\phi$-Lagrangian associated with Eq.~\ref{casdecaction} is the analogue of the $\pi$-Lagrangian in DGP gravity~\cite{luty, nicolis}. It should capture  local physics on the   3-brane, corresponding to normalisable fluctuations in the $5D$ bulk. 
Note that this means it cannot capture the self tuning behaviour of the full $6D$ theory.  This is because the self tuning solutions correspond to  non-local, non-normalisable fluctuations about the vacuum of the effective $5D$ theory described by the cascading cosmology model (\ref{cascact}). We could, of course, consider normalisable fluctuations about a non-trivial self tuning solution. We expect that this would also give a bi-galileon theory in the appropriate decoupling limit. For sufficiently large 3-brane tension it would actually correspond to a ghost-free bi-galileon theory \cite{casc1}, unlike the example given here. However, a detailed analysis is outside of the scope of this paper.

\section{Bi-galileon modification of gravity} \label{sec:modgrav}
The cascading cosmology model described in the previous section is an example of a modification of General Relativity by two additional galileon fields, $\pi$ and $\phi$. In a generic theory of modified gravity, the amplitude for the exchange of one graviton between two conserved sources, $T_{\mu\nu}$ and $T_{\mu\nu}'$, is given by
\be
{\cal A}=T_{\mu\nu}D^{\mu\nu \alpha \beta} T_{\alpha \beta}'\,,
\ee
where $D^{\mu\nu \alpha \beta}$ is the graviton propagator. Now, for General Relativity, the amplitude is given by
\be
{\cal A}_{GR}=-\frac{2}{M_{pl}^2} \left(T^{\mu\nu}\frac{1}{\square_4} T_{\mu\nu}'-\frac{1}{2}T\frac{1}{\square_4} T'\right)\,,
\ee
where $T=T^\mu{}_\mu$. We will be interested in the case where  gravity gets modified by two additional scalars so that locally we have
\be
\delta {\cal A}={\cal A}-{\cal A}_{GR}= T \frac{1}{\alpha_1 \square_4}  T'+ T \frac{1}{\alpha_2\square_4}  T'\,.
\ee
Such a theory can be described by the following action
\be
S=\int d^4 x~ \frac{M_{pl}^2}{2}\sqrt{-\tilde g} R(\tilde g)-\frac{\alpha_1}{2}\pi_1 \square_4 \pi_1-\frac{\alpha_2}{2}\pi_2 \square_4 \pi_2 +\frac{1}{2} \tilde h_{\mu\nu} T^{\mu\nu}+\pi_1 T+\pi_2 T +\text{interactions}\,,
\ee
where $\tilde g_{\mu\nu}=\eta_{\mu\nu}+\tilde h_{\mu\nu}$. The  fluctuation $\tilde h_{\mu\nu}$ is identified with the graviton in General Relativity since it satisfies the same linearised field equations. Therefore, for a given source and boundary conditions, the solution $\tilde h_{\mu\nu}$ coincides with the linearised GR solution. This statement is true to all orders in the limit,
\be
M_{pl},\; \alpha_i,\; T^{\mu\nu} \to \infty\,,\qquad  \frac{\alpha_i}{M_{pl}^2}=\text{const}\,, ~\frac{T^{\mu\nu}}{M_{pl}}=\text{const}\,, \label{limit}
\ee
where the scalars decouple from $\tilde h_{\mu\nu}$. The physical graviton differs from its GR counterpart by a scalar fluctuation
\be
h_{\mu\nu}=\tilde h_{\mu\nu}+2(\pi_1+\pi_2)\eta_{\mu\nu} \,,\label{physh}
\ee
since the physical metric seen by matter is given by $g_{\mu\nu}=\eta_{\mu\nu}+h_{\mu\nu}$. 

Let us consider the decoupling limit (\ref{limit}), with the additional assumption that the strength of all scalar interactions are held fixed. This basically means we are neglecting the backreaction of the scalars on the geometry, so we can consider them as fields on Minkowski space. Note that we retain scalar self-interactions because they are essential for  bi-galileon corrections  to GR to be extrapolated from $O(10^{-5})$ in the solar system to $O(1)$ at  cosmological scales.  Significant modifications to gravity at Hubble distances would inevitably have a role to play in the dark energy problem.  The action is given by
\be
S=\int d^4 x~ -\frac{M_{pl}^2}{4}\tilde h^{\mu\nu}{\cal E} \tilde h_{\mu\nu}+\frac{1}{2} \tilde h_{\mu\nu} T^{\mu\nu}+\pi_1 T+\pi_2 T +{\cal L}_{\pi_1,\pi_2}\,,
\ee
where ${\cal E}\tilde h_{\mu\nu}=-\frac{1}{2} \square_4 \left(\tilde h_{\mu\nu}-\frac{1}{2} \tilde h \eta_{\mu\nu}\right)+\ldots$ is the linearised Einstein tensor, and the Lagrangian ${\cal L}_{\pi_1, \pi_2}$ only depends on the scalars $\pi_1$ and $\pi_2$.  As we can see from Eq.~\ref{deccascact}, the action for the  cascading cosmology certainly takes this form in the decoupling limit.  In Eq.~\ref{deccascact} the scalar action is Galilean invariant. To generalise this, we are interested in bi-galileon modifications of gravity for which ${\cal L}_{\pi_1, \pi_2}$ is the Lagrangian for a bi-galileon theory. This means that if $\pi_i \to \pi_i+(b_i)_\mu x^\mu+c_i$ then ${\cal L}_{\pi_1, \pi_2} \to {\cal L}_{\pi_1, \pi_2} +\text{total derivative}$.

Given the form of the physical graviton (\ref{physh}), it is convenient to define new galileon fields $\pi=\pi_1+\pi_2$ and $\xi=\pi_1-\pi_2$, so that only one of the scalars couples to matter.  The action now takes the form
\be \label{act}
S=\int d^4 x~ -\frac{M_{pl}^2}{4}\tilde h^{\mu\nu}{\cal E} \tilde h_{\mu\nu}+\frac{1}{2} \tilde h_{\mu\nu} T^{\mu\nu}+\pi T +{\cal L}_{\pi,\xi}\,,
\ee
where ${\cal L}_{\pi,\xi}$ corresponds to a bi-galileon theory. The physical graviton is now
\be \label{galhrel}
h_{\mu\nu}=\tilde h_{\mu\nu}+2\pi \eta_{\mu\nu}\,.
\ee
Although only $\pi$ couples to matter directly, $\xi$ couples to matter {\it in}directly, through its mixing with $\pi$.  It is possible for the two scalars to be mixed at quadratic order, as well as at higher order. 

Given that we have taken all our fields to propagate on Minkowski space, one might be forgiven for thinking that we cannot say anything about cosmology, or perturbations about cosmological solutions, including de Sitter vacua.  Fortunately, this is not the case. We can think of any spatial flat cosmological spacetime as a local perturbation around Minkowski space~\cite{gal}. In this context, {\it local} means local in both space and time, at sub-horizon distances and over a sub-Hubble time. If we take {\it here} to be $\vec x=0$ and {\it now} to be $t=0$, then for $|\vec x| \ll H^{-1}$ and $|t| \ll H^{-1}$,  we have~\cite{gal}
\be
ds^2=-d\tau^2+a(\tau)^2 d\vec y^2 \approx \left[ 1-\frac{1}{2}H^2 |\vec x|^2+\frac{1}{2}(2\dot H +H^2)t^2\right](-dt^2+d\vec x^2)\,,
\ee
where the Hubble scale $H$ and its time derivative $\dot H$ may be evaluated at $t=0$.  This is a perturbation about Minkowski space written in Newtonian gauge
\be
ds^2 \approx -(1+2\Phi)dt^2+(1-2\Psi)d\vec x^2\,,
\ee
where the Newtonian potentials are
\be
\Phi=-\frac{1}{4}H^2 |\vec x|^2+\frac{1}{4}(2\dot H +H^2)t^2\,, \qquad \Psi=-\Phi\,.
\ee
For a given cosmological fluid, the corresponding GR solutions have Hubble parameter $H_{GR}$. Since $\tilde h_{\mu\nu}$ agrees with the linearised GR solution, we have $\tilde h_{tt}=-2\Phi_{GR}, \tilde h_{ij}=2\Phi_{GR}\delta_{ij}$, where
 \be
\Phi_{GR}=-\frac{1}{4}H_{GR}^2 |\vec x|^2+\frac{1}{4}(2\dot H_{GR} +H_{GR}^2)t^2\,.
\ee
Now in our modified theory  the physical Hubble parameter (associated with $h_{\mu\nu}$) differs from the corresponding GR value, $H \neq H_{GR}$. Due to Eq.~\ref{galhrel}, we have a non-trivial scalar
\be
\pi= \Phi-\Phi_{GR}\,.
\ee
The other scalar, $\xi$,  follows from the equations of motion. Note that a Galilean transformation $\pi \to \pi+b_\mu x^\mu+c$ merely corresponds to a coordinate transformation $x^\mu\to x^\mu-cx^\mu+\frac{1}{2}(x_\nu x^\nu b^\mu-2b_\nu x^\nu x^\mu)$ in the physical metric. 

Of particular interest are maximally symmetric vacua, for which $H$ and $H_{GR}$ are constant. Indeed, if the vacuum energy is $\sigma$, we have $H_{GR}^2=\sigma/3M_{pl}^2$. It is easy to check that one scalar now takes the particularly simple form $\bar \pi=-\frac{1}{4} k_\pi x_\mu x^\mu$ where $k_\pi=H^2-H^2_{GR}$.  In principle the other scalar can be anything, but we will typically assume that it takes the same form on maximally symmetric solutions, so that $\bar \xi=-\frac{1}{4} k_\xi x_\mu x^\mu$.

Ultimately, we would like to promote our galileon theory to a fully covariant theory of gravity, coupled to a pair of scalar fields. Whilst this will inevitably break the Galilean invariance \cite{covgal}, we should expect to recover the galileon description in the decoupling limit $M_{pl} \to \infty$. One does have to be slightly careful not to introduce any higher derivatives by accident when covariantising the theory, but this can be done using additional non-minimal couplings \cite{covgal}. As a step towards this, let us first  transform the action (\ref{act}) to ``Jordan" frame by rewriting everything in terms of the physical metric (\ref{galhrel}). This gives   
\be \label{jordan}
S=\int d^4 x~ -\frac{M_{pl}^2}{4}h^{\mu\nu}{\cal E}  h_{\mu\nu}-M_{pl}^2 \pi (\pd_\mu\pd_\nu h^{\mu\nu}-\pd^2 h)+\frac{1}{2} h_{\mu\nu} T^{\mu\nu}+\hat {\cal L}_{\pi,\xi},
\ee
where $\hat {\cal L}_{\pi, \xi}={\cal L}_{\pi,\xi}-3M_{pl}^2\pi \pd^2 \pi $. The covariant completion of this theory will take the form
\be \label{covact}
S_{cov}=\int d^4 x \sqrt{-g}~\left[ \frac{M_{pl}^2}{2}(1-2\pi)R+{\cal L}_\text{matter}+ {\cal L}_\text{scalar}\right]
\ee
where ${\cal L}_\text{matter}[g; \Psi_n]$ is the matter Lagrangian, describing matter fields, $\Psi_n$, minimally coupled to gravity. The scalar contribution, $ {\cal L}_\text{scalar}[g; \pi, \xi]$, is constructed as follows: take $\hat {\cal L}_{\pi,\xi}$ and let $\eta_{\mu\nu} \to g_{\mu\nu}, ~\pd_\mu \to \nabla_\mu$, then add non-minimal couplings to curvature to compensate for the higher derivative terms in the equations of motion. 

Actually, this completion is useful in that it enables us to establish the validity of our galileon description. Recall that this neglects the backreaction of the scalars onto the geometry. From Eq. \ref{covact}, we see that the scalars contribute an energy-momentum tensor
\be
T_\text{scalar}^{\mu\nu}[g; \pi, \xi]=\frac{2}{\sqrt{-g}} \frac{\delta }{\delta g_{\mu\nu}}\int d^4 x \sqrt{-g}~{\cal L}_\text{scalar}
\ee
The gravitational field equations arising from the covariant action (\ref{covact}) are given by
\be \label{graveq}
\frac{M_{pl}^2}{2} (1-2 \pi) G^{\mu\nu}+M_{pl}^2 (\nabla^\mu\nabla^\nu -g^{\mu\nu} \Box)\pi-\frac{1}{2}\left( T^{\mu\nu}+T_\text{scalar}^{\mu\nu}\right)=0
\ee
Although the galileon description includes non-linear effects at the level of the scalar field equations, it neglects them at the level of the gravity equation. Expanding Eq. \ref{graveq} about Minkowski space, $g_{\mu\nu}=\eta_{\mu\nu}+h_{\mu\nu}$, we find that
\be \label{lingraveqn}
\frac{M_{pl}^2}{2}  {\cal E} h^{\mu\nu}+M_{pl}^2 \left(\pd^\mu \pd^\nu -\eta^{\mu\nu} \pd^2\right)\pi-\frac{1}{2}T^{\mu\nu}[\eta; \Psi_n]=\frac{1}{2} T_\text{scalar}^{\mu\nu}[\eta; \pi, \xi]+ \ldots
\ee
where ``$\ldots$" denotes terms that are ${\cal O}(h)$ suppressed relative to the terms on the LHS of Eq. \ref{lingraveqn}. The gravity equation in the galileon description corresponds to neglecting everything on the RHS of Eq. \ref{lingraveqn}.  When is this justified? For a small graviton fluctuation, we can consistently neglect those  the terms that are ${\cal O}(h)$ suppressed, denoted by ``$\ldots$".  However, even then, the backreaction of the scalars on to the geometry can be significant.  In other words, it is not clear that the energy momentum of the scalars, evaluated on Minkowski,   can be consistently neglected. To justify neglecting this backreaction  we conservatively require that $T_\text{scalar}^{\mu\nu} [\eta,;\pi, \xi] \ll M_{pl}^2   {\cal E} h^{\mu\nu} $. 

Provided this condition holds, we will see that our bi-galileon theory allows for a rich and interesting phenomenology, to be studied in detail in our companion paper~\cite{otherpaper}. As we have just discussed, we can describe cosmological solutions, such as de Sitter space, as local perturbations about Minkowski.   We will also be interested in fluctuations about these solutions, and in particular about non-trivial maximally symmetric vacua. These just correspond to fluctuations in the galileon fields about their background values. As we mentioned earlier,  the non-linearity of the galileon action will be essential to the success of escaping the local gravity tests and producing interesting cosmological effects simultaneously.
\section{Constructing the bi-galileon theory} \label{sec:gentheory}
Let us now construct the most general bi-galileon theory. Our task is to find the most general Lagrangian describing two scalars, $\mathcal{L}_{\pi,\xi}$, that is invariant under the Galilean transformation
\be
\pi\to \pi+b_{\mu}x^{\mu}+c\,,\qquad \xi\to \xi+\tilde b_{\mu}x^{\mu}+\tilde c\,,
\ee
This amounts to requiring that  the corresponding equations of motion  are built exclusively out of second derivative terms $\pd_\mu\pd_\nu\pi$, $\pd_\alpha\pd_\beta\xi$ contracted with the background Minkowski metric . Single, or zero derivative contributions to the equations of motion break the Galilean  invariance, whereas  higher derivatives introduce extra degrees of freedom and ghosts. 

The Galilean symmetry greatly constrains the structure of the Lagrangian. For example, under an infinitesimal constant shift $\pi\to \pi+c$, the Lagrangian $\mathcal{L}_{\pi,\xi}$ varies as
\be
\mathcal{L}_{\pi,\xi} \to \mathcal{L}_{\pi,\xi}
+\frac{\pd\mathcal{L}_{\pi,\xi}}{\pd\pi}c\,.
\ee
However, as this shift happens to be a Galilean transformation, we know that any change in the Lagrangian must be a total derivative. It follows that  $\pd\mathcal{L}_{\pi,\xi}/\pd\pi=\pd_\mu \mathcal{J}^\mu$, for some ${\cal J}^\mu$. Now the $\pi$ equation of motion goes like 
\be 
\frac{\delta }{\delta \pi}\int  d^4 x~{\cal L}_{\pi, \xi}=\frac{\pd\mathcal{L}_{\pi,\xi}}{\pd\pi}-\pd_\mu\left[\frac{\pd\mathcal{L}_{\pi,\xi}}{\pd(\pd_\mu\pi)}-\ldots\right]
\ee
which is clearly a total derivative $\pd_\mu\left[ {\cal J}^\mu- \pd\mathcal{L}_{\pi,\xi}/\pd(\pd_\mu\pi)+\ldots\right]$. Indeed, the $\pi$ equation of motion is a total derivative at each order in $\pi$ and $\xi$. We can apply a similar argument to the $\xi$ equation of motion. 

We can now formulate the problem as follows: at $(m,n)$-th order ($m$-th order in $\pi$ and $n$-th order in $\xi$), what is the most general total derivative built out of second derivative terms? One obvious total derivative is given by
\be \label{eommn}
\mathcal{E}_{m,n}= (m+n)! \delta^{\mu_1}_ {[\nu_1} \ldots \delta^{\mu_m} _{\nu_m}
             ~\delta^{\rho_1} _{\sigma_1} \ldots \delta^{\rho_n}_{ \sigma_n]}
           (\pd_{\mu_1}\pd^{\nu_1}\pi) \ldots  (\pd_{\mu_m}\pd^{\nu_m}\pi)
             ~(\pd_{\rho_1}\pd^{\sigma_1}\xi) \ldots  (\pd_{\rho_n}\pd^{\sigma_n}\xi)\,.
\ee
It is easy to see all such terms with $m+n>4$  vanish in four dimensions. We will now write out all the non-vanishing  $\mathcal{E}_{m,n}$ explicitly in 4D. For notational convenience, we suppress the Lorentz indices and separate Lorentz scalars by `$\cdot$' when it is confusing. For example, we have $\pd\pd\pi\pd\pd\xi \cdot \pd\pd\pi\pd\pd\xi \equiv (\pd_\mu \pd_\nu \pi)( \pd^\nu \pd^\mu \xi) (\pd_\alpha\pd_\beta\pi)( \pd^\beta\pd^\alpha\xi)$. For $m \geq n$, the non-vanishing terms are
\bea
\mathcal{E}_{0,0}     &=&   1\,,  \nn
\mathcal{E}_{1,0}     &=&   \Box\pi\,,   \nn
\mathcal{E}_{2,0}     &=&   (\Box\pi)^2-(\pd\pd\pi)^2\,,   \nn
\mathcal{E}_{1,1}     &=&   \Box\pi \Box\xi-\pd\pd\pi \pd\pd\xi\,,   \nn
\mathcal{E}_{3,0}     &=&   (\Box\pi)^3-3\Box\pi  (\pd\pd\pi)^2+2(\pd\pd\pi)^3\,,  \nn
\mathcal{E}_{2,1}     &=&   (\Box\pi)^2 \Box\xi
                           -\Box\xi(\pd\pd\pi)^2-2\Box\pi\pd\pd\pi\pd\pd\xi
                           +2(\pd\pd\pi)^2\pd\pd\xi\,,   \nn
\mathcal{E}_{4,0}     &=&   (\Box\pi)^4-6(\Box\pi)^2(\pd\pd\pi)^2+8\Box\pi(\pd\pd\pi)^3
                           +3(\pd\pd\pi)^2\cdot(\pd\pd\pi)^2-6(\pd\pd\pi)^4\,,   \nn
\mathcal{E}_{3,1}     &=&   (\Box\pi)^3\Box\xi-3\Box\pi\Box\xi(\pd\pd\pi)^2
                           -3(\Box\pi)^2\pd\pd\pi\pd\pd\xi+2\Box\xi(\pd\pd\pi)^3  \nn
           &\phantom{=}&   +6\Box\pi(\pd\pd\pi)^2\pd\pd\xi+3(\pd\pd\pi)^2\cdot\pd\pd\pi\pd\pd\xi
                           -6(\pd\pd\pi)^3\pd\pd\xi\,,     \nn
\mathcal{E}_{2,2}     &=&   (\Box\pi)^2(\Box\xi)^2-(\Box\xi)^2(\pd\pd\pi)^2-(\Box\pi)^2(\pd\pd\xi)^2
                           -4\Box\pi\Box\xi\pd\pd\pi\pd\pd\xi    \nn
           &\phantom{=}&   +4\Box\xi(\pd\pd\pi)^2\pd\pd\xi+4\Box\pi\pd\pd\pi(\pd\pd\xi)^2
                           +(\pd\pd\pi)^2\cdot(\pd\pd\xi)^2  \nn
           &\phantom{=}&   +2\pd\pd\pi\pd\pd\xi\cdot\pd\pd\pi\pd\pd\xi-4(\pd\pd\pi)^2(\pd\pd\xi)^2
                           -2(\pd\pd\pi\pd\pd\xi)^2\,,
\eea
For $m < n$, we take $\mathcal{E}_{m,n}=\mathcal{E}_{n,m}|_{\pi\leftrightarrow\xi}$. Note that in each case ${\cal E}_{m,n}=(\Box \pi)^m(\Box\xi)^n+\ldots$.

We will now prove that Eq.~\ref{eommn} is the {\it only} total derivative of order $(m,n)$ that is second order in spacetime derivatives. The proof works along the same lines as the corresponding single galileon one given in~\cite{gal}. First, assume there is another suitable total derivative $\mathcal{E}'_{m,n}$. Now consider $\mathcal{E}'_{m,n}$ as a Lagrangian density. As  $\mathcal{E}'_{m,n}$ is a total derivative, the  resulting $\pi$ and $\xi$ equations of motion must be trivial
\be \label{triv}
\frac{\delta }{\delta \pi}\int  d^4 x ~\mathcal{E}'_{m,n}=0\,, \qquad \frac{\delta }{\delta \xi}\int  d^4 x ~\mathcal{E}'_{m,n}=0\,.
\ee
Without loss of generality, a general term in $\mathcal{E}'_{m,n}$ goes like\footnote{Note that we have suppressed permutations of $\pi$ and $\xi$ within Lorentz scalars; for instance, we collectively list the Lorentz scalars $\pd \pd \pi\pd\pd\xi\pd \pd \pi\pd\pd\xi$ and $(\pd \pd \pi)^2(\pd\pd\xi)^2$ as $(\pd \pd \pi)^2(\pd\pd\xi)^2$} 
\be\label{eq:term}
(\Box \pi)^{i_0}(\Box \xi)^{j_0}(\pd\pd \pi)^{i_1}(\pd\pd \xi)^{j_1}\cdot\ldots\cdot(\pd\pd \pi)^{i_k} (\pd\pd \xi)^{j_k}\,,
\ee 
where $ i_0+\ldots+i_k=m, ~j_0+\ldots+j_k=n$. Let us now consider its contribution to the resulting  equations of motion (\ref{triv}). A term in the $\pi$ equations of motion  can be obtained by shifting $\pd\pd$ (along with the Lorentz indices) from one $\pd\pd\pi$ to another $\pd\pd\pi$ or another $\pd\pd\xi$ and then dropping the $\pi$ without derivatives. Therefore Eq.~\ref{eq:term}  must contribute a term proportional to 
\be \label{proeomterm}
(\Box \pi)^{i_0}(\Box \xi)^{j_0}\Box\pd\pd \pi(\pd\pd \pi)^{i_1-2}(\pd\pd \xi)^{j_1}\cdot\ldots\cdot(\pd\pd \pi)^{i_k} (\pd\pd \xi)^{j_k}\,,
\ee
in the equation of motion. As the overall equation of motion is trivial,   this term must be cancelled by the variation of another term in $\mathcal{E}'_{m,n}$. Aside from Eq.~\ref{eq:term}, there is only one other possible term that can produce a term like Eq.~\ref{proeomterm} in the $\pi$ equation of motion. It is given by
\be
(\Box \pi)^{i_0+1}(\Box \xi)^{j_0}(\pd\pd \pi)^{i_1-1}(\pd\pd \xi)^{j_1}\cdot\ldots\cdot(\pd\pd \pi)^{i_k} (\pd\pd \xi)^{j_k}\,.
\ee
It follows that this must also be in $\mathcal{E}'_{m,n}$. One can easily imagine a similar argument applies to $\xi$. Now, if we apply these arguments iteratively, we would conclude that a term proportional to $(\Box \pi)^m(\Box \xi)^n$ must be in $\mathcal{E}'_{m,n}$. Then we immediately run into a contradiction, because linearly combining $\mathcal{E}_{m,n}$ and $\mathcal{E}'_{m,n}$ should produce another total derivative which is free of $(\Box \pi)^m(\Box \xi)^n$. This proves that the only suitable total derivative at order $(m, n)$ is $\mathcal{E}_{m,n}$.

Returning to the full theory, the equations of motion for $\pi$ and $\xi$ in general correspond to linear combinations of the ${\cal E}_{m, n}$:
\bea
 \frac{\delta }{\delta \pi}\int  d^4 x~{\cal L}_{\pi, \xi}=\sum_{0 \leqslant m+n\leqslant 4} a_{m,n}\mathcal{E}_{m,n}\,,\nn
  \frac{\delta }{\delta \xi}\int  d^4 x~{\cal L}_{\pi, \xi}=\sum_{0 \leqslant m+n\leqslant 4} b_{m,n}\mathcal{E}_{m,n}\,.
\eea
Since these equations are derived from the same Lagrangian, ${\cal L}_{\pi, \xi}$, we anticipate  an integrability condition relating the coefficients $a_{m, n}$ and $b_{m, n}$. To construct the Lagrangian, we note that equations of motion with the correct form can be derived from the following
\be \label{lagpixi}
\mathcal{L}_{\pi,\xi} = \sum_{m+n\leqslant 4}\mathcal{L}_{m,n}\,,
\ee
where
\be \label{Lmn}
\mathcal{L}_{m,n} = (\alpha_{m,n}\pi+\beta_{m,n}\xi){\cal E}_{m,n}\,.
\ee
To see this note that
\bea
 \frac{\delta }{\delta \pi}\int  d^4 x~{\cal L}_{m, n} &=&(m+1)\alpha_{m,n}\mathcal{E}_{m,n}+m\beta_{m,n}\mathcal{E}_{m-1,n+1}\,, \nn
 \frac{\delta }{\delta \xi}\int  d^4 x~{\cal L}_{m, n} &=&n\alpha_{m,n}\mathcal{E}_{m+1,n-1}+(n+1)\beta_{m,n}\mathcal{E}_{m,n}\,, \nonumber
 \eea
which yields the following
\begin{align}
& \frac{\delta }{\delta \pi}\int  d^4 x~{\cal L}_{\pi, \xi}=\sum_{0 \leqslant m+n\leqslant 4 }(m+1)(\alpha_{m,n}+\beta_{m+1,n-1})\mathcal{E}_{m,n} & (\beta_{i,-1}=0)\,,\\
& \frac{\delta }{\delta \xi}\int  d^4 x~{\cal L}_{\pi, \xi}=\sum_{0 \leqslant m+n\leqslant 4 }(n+1)(\beta_{m,n}+\alpha_{m-1,n+1})\mathcal{E}_{m,n} & (\alpha_{-1,i}=0)\,.
\end{align}
Thus we have the relations between the coefficients of the Lagrangian and the equations of motion:
\begin{align}\label{coab1}
&a_{m,n}=(m+1)(\alpha_{m,n}+\beta_{m+1,n-1})\,,\\
\label{coab2}
&b_{m,n}=(n+1)(\beta_{m,n}+\alpha_{m-1,n+1})\,.
\end{align}
As expected there exists an integrability condition relating the coefficients in the equation of motion. One can easily check that 
\be
na_{m-1, n}-mb_{m, n-1}=0\,. \label{abrel}
\ee
The origin of this condition lies in the fact that $\pi {\cal E}_{m-1, n}-\xi {\cal E}_{m, n-1}$ is a total derivative for $n, m \geq 1$. This means we have included a number of total derivatives in the Lagrangian that do not contribute to the dynamics. Indeed, at order $m+n+1 \geq 2$ the total number of free parameters is not $2(m+n+1)$ as one might naively expect, but $m+n+2$ owing to the constraints given by Eq. \ref{abrel}.

We now have the most general bi-galileon theory, given by Eqs.~\ref{lagpixi}, \ref{Lmn} and \ref{eommn}. Referring back to the discussion in the previous section, we note that  only $\pi$ couples directly to matter,  through an interaction $\int d^4 x~\pi T$. It follows that the equations of motion are given by
\bea
T+\sum_{0 \leqslant m+n\leqslant 4} a_{m,n}\mathcal{E}_{m,n}&=&0\,, \label{eompi} \\
\sum_{0 \leqslant m+n\leqslant 4} b_{m,n}\mathcal{E}_{m,n}&=&0\,, \label{eomxi}
\eea
where the $a_{m, n}$ and the $b_{m,n}$ are related according to Eq.~\ref{abrel}.

We conclude this section by stating, without proof, the generalisation to $N$ galileons. The analogue of ${\cal E}_{m, n}$ is
\bea
 {\cal E}_{m_1, \ldots, m_N}&=& \left(m_1+\ldots+ m_N\right) ! \delta^{\mu_1}_ {[\nu_1} \ldots \delta^{\mu_{m_1}} _{\nu_{m_1}}\ldots
             \delta^{\rho_1} _{\sigma_1} \ldots \delta^{\rho_{m_N}}_{ \sigma_{m_N}]} \nn
        &&\qquad \qquad\times~  \left[ (\pd_{\mu_1}\pd^{\nu_1}\pi_1) \ldots  (\pd_{\mu_{m_1}}\pd^{\nu_{m_1}}\pi_1)\right] \ldots 
          \left[
           (\pd_{\rho_1}\pd^{\sigma_1}\pi_N) \ldots  (\pd_{\rho_{m_N}}\pd^{\sigma_{m_N}}\pi_N)\right]\,.
\eea
Therefore, the general $N$-galileon Lagrangian is given by
\be
\mathcal{L}_{\pi_1, \ldots, \pi_N} = \sum_{m_1+\ldots+m_N\leqslant 4}\mathcal{L}_{m_1,\ldots,m_N}\,,
\ee
where
\be
\mathcal{L}_{m_1, \ldots, m_N}=(\alpha_{m_1, \ldots, m_N}^1\pi_1+\ldots+\alpha^N_{m_1, \ldots, m_N}\pi_N){\cal E}_{m_1, \ldots, m_N}\,.
\ee

\section{Maximally symmetric vacua and their stability}\label{sec:maxsym}
 Although a detailed study of bi-galileon phenomenology will be left to our companion paper~\cite{otherpaper},  we will now initiate a study of maximally symmetric vacua. We will be particularly interested in establishing stability conditions associated with ghost-like excitations of the galileon fields about these vacua. 
 
 We will allow for a non-zero vacuum energy so that the background stress energy tensor is given by $\bar T^\mu_\nu=-\sigma \delta^\mu_\nu$. As discussed in section \ref{sec:modgrav}, a maximally symmetric solution has a background $\pi$ field,
 \be \label{pibg}
\bar \pi(x)=-\frac{1}{4}k_\pi x_\mu x^\mu\,, ~~~~ \pd_\mu \pd_\nu \bar \pi=-\frac{1}{2}k_\pi
 \eta_{\mu\nu} \,,
\ee
 where $k_\pi=H^2-H^2_{GR}$ is the difference between the Hubble parameters calculated in the modified gravity theory and in GR. For simplicity we assume that the background $\xi$ field takes the same form
\be  \label{xibg}
\bar \xi(x)=-\frac{1}{4}k_\xi x_\mu x^\mu\,, \qquad \pd_\mu \pd_\nu \bar \xi=-\frac{1}{2}k_\xi \eta_{\mu\nu}\,,
\ee
for some constant $k_\xi$. If our theory happens to be derived from a brane model, we expect this choice to correspond to choosing maximal symmetry in the bulk and on the branes.  We now plug this into our field equations Eqs.~\ref{eompi} and \ref{eomxi}. This yields
\bea
\label{bg1}
E_\pi(k_\pi, k_\xi)&\equiv&-4\sigma+\sum_{0 \leqslant m+n\leqslant 4} \left(-\frac{1}{2}\right)^{m+n} \frac{4!}{(4-m-n)!}a_{m, n}k_\pi^m k_\xi^n 
=0\,,\\
\label{bg2}
E_\xi(k_\pi, k_\xi)&\equiv&\sum_{0 \leqslant m+n\leqslant 4} \left(-\frac{1}{2}\right)^{m+n} \frac{4!}{(4-m-n)!}b_{m, n}k_\pi^m k_\xi^n  =  0\,,
\eea
where we recall that $a_{m, n}$ and $b_{m, n}$ are given by Eqs.~\ref{coab1} and \ref{coab2}. Generically these represent two independent equations in two unknowns. Being quartic polynomials, their solutions  will typically contain sixteen  distinct points in the real $(k_\pi, k_\xi)$ plane. Of course, there may be degenerate configurations with a continuum of solutions, or other configurations with no real solutions whatsoever. We will study these aspects in more detail in \cite{otherpaper}. For the moment, let us assume that  a real solution exists.

Can we really trust our solution? We can think of two things in particular that could spoil the background: gravitational back reaction and/or dynamical instabilities. The former refers to backreaction of the scalars onto the geometry whereas the latter refers to ghost-like excitations. To succinctly express suitable conditions on the background,  it is convenient to consider what happens when we plug our background fields directly into the action. It turns out that
\be
\int d^4 x ~  \left({\cal L}_{\bar \pi, \bar \xi}+T\pi\right)=\frac{1}{4} \left(\int d^4 x~x_\mu x^\mu\right)L(k_\pi, k_\xi)\,,
\ee
where we define the {\it action polynomial},
\be
L(k_\pi, k_\xi)=4\sigma k_\pi-\sum_{0 \leqslant m+n\leqslant 4}(\alpha_{m,n}k_\pi+\beta_{m, n} k_\xi) \left(-\frac{1}{2}\right)^{m+n} \frac{4!}{(4-m-n)!}k_\pi^m k_\xi^n  \,.
\ee
Now, it is easy to check, using Eqs.~\ref{coab1} and \ref{coab2}, that
\be
E_\pi(k_\pi, k_\xi)=-\frac{\pd L}{\pd k_\pi}\,, \qquad E_\xi(k_\pi, k_\xi)=-\frac{\pd L}{\pd k_\xi}\,. \label{E=pdL}
\ee
Of course this comes as no surprise since the equations of motion are designed to extremise the action. For the maximal symmetry case in question this  simply corresponds to extremising $L$. 

Now let us return to the issue of backreaction of the scalars on to the geometry. This was discussed at the end of section \ref{sec:modgrav}. Schematically, we expect
\be
T^{\mu\nu}_\text{scalar}[\eta; \bar \pi, \bar \xi] =
x^2 \sum_{0 \leqslant m+n \leqslant 4} {\cal O}(1)(\alpha_{m,n}-3M_{pl}^2 \delta_m^1 \delta_n^0 )k_\pi^{m+1} k_\xi^n  +{\cal O}(1) \beta_{m,n} k_\pi^{m} k_\xi^{n+1} 
\ee
 Recall that to justify neglecting the backreaction, we require $T_\text{scalar}^{\mu\nu}  \ll M_{pl}^2 {\cal E} h_{\mu\nu} \sim  M_{pl}^2 H^2$.
 
Assuming that we can indeed neglect the backreaction as described, we then need to ask if our background solution is stable against excitation of ghosts? To study the stability of a solution we need to examine its fluctuations, so we shift the fields so that they represent fluctuations on the appropriate background
\be  \label{shift}
\pi\to \bar \pi+\pi\,, \qquad \xi\to \bar \xi+\xi\,.
\ee
We claim that the background solution has a  ghost-like instability if the Hessian of $L(k_\pi, k_\xi)$
\be
Hess(L)=\left(\begin {array}{cc}  \frac{\pd^2 L}{\pd k_\pi^2} &\frac{\pd^2 L}{\pd k_\pi \pd k_\xi} \\
\frac{\pd^2 L}{\pd k_\xi \pd k_\pi} &\frac{\pd^2 L}{\pd k_\xi^2} \end{array}\right)
\ee
has a {\it negative} eigenvalue on the solution. Let us  now prove this claim.

The field re-definition (\ref{shift}) takes us from the trivial to a non-trivial background, and gives rise to the following shift in the  Lagrangian
\be
\mathcal{L}_{\pi,\xi} \to { \mathcal{L}_{\bar \pi,\bar \xi}} + {\mathcal{L}^{\prime}_{\pi,\xi}}\,.
\ee
It is clear that the perturbed Lagrangian ${\mathcal{L}^{\prime}_{\pi,\xi}}$ has the same symmetries as $\mathcal{L}_{\pi,\xi}$, so we may once again write
\be \label{lagpixiP}
\mathcal{L}^\prime_{\pi,\xi} = \sum_{1\leqslant m+n\leqslant 4}\mathcal{L}^\prime_{m,n}\,,
\ee
where
\be
\mathcal{L}^\prime_{m,n} = (\alpha'_{m,n}\pi+\beta'_{m,n}\xi){\cal E}_{m, n}\,.
\ee
Note that, due to the background equations of motion, linear terms in $\mathcal{L}^\prime_{\pi,\xi}$ vanish; that is there is no contribution from $m=n=0$. The equations of motion are given by 
\bea
&&\sum_{1 \leqslant m+n\leqslant 4} a'_{m,n}\mathcal{E}_{m,n}=-\delta T\,, \label{eompids} \\
&&\sum_{1 \leqslant m+n\leqslant 4} b'_{m,n}\mathcal{E}_{m,n}=0\,, \label{eomxids}
\eea
with the relations
\begin{align}\label{cocd1}
a'_{m,n}&=(m+1)(\alpha'_{m,n}+\beta'_{m+1,n-1})&(\beta'_{i,-1}=0) \,,\\
\label{cocd2}
b'_{m,n}&=(n+1)(\beta'_{m,n}+\alpha'_{m-1,n+1})&(\alpha'_{-1,i}=0)\,,
\end{align}
and
\be
na'_{m-1, n}-mb'_{m, n-1}=0\,. \label{cdrel}
\ee
There is a linear map between the parameters in our original theory $a_{m,n}, ~b_{m,n}$ and the perturbed theory $a'_{m,n},~b'_{m,n}$.  By direct substitution we find that for $1 \leqslant m+n \leqslant 4$, we have
\be
a'_{m, n}=\sum_{i=m}^4\sum_{j=n}^4 M_{m, n}{}^{i, j}a_{i, j}\,, \qquad b'_{m, n}=\sum_{i=m}^4\sum_{j=n}^4 M_{m, n}{}^{i, j}b_{i,j}\,, \label{map}
\ee
where
\be
M_{m, n}{}^{i, j}=\left(-\frac12\right)^{i+j-m-n} \left( \begin{array}{c} i \\ m\end{array}\right) \left( \begin{array}{c} j \\ n\end{array}\right) \frac{(4-m-n)!}{(4-i-j)!}k_\pi^{i-m}k_\xi^{j-n}\,. 
\ee
Here the combination $\left( \begin{array}{c} i \\ m\end{array}\right)=i!/m!(i-m)!$. For $N<0$, we extend the definition of ``factorial" using the Gamma function, $N!=\Gamma(N+1)$, and recall that $1/\Gamma(N+1)=0$ when $N$ is a negative integer. It immediately follows that we only get contributions from $i+j \leqslant 4$.

To study the stability of the background, we consider the leading order piece of the perturbed Lagrangian
\bea
{\cal L}'_{\pi, \xi}&=&(\alpha'_{10} \pi+\beta'_{10} \xi)\Box \pi+(\alpha'_{01} \pi+\beta'_{01} \xi)\Box \xi+\ldots \nn
&=&-\left[\alpha'_{10}(\pd \pi)^2+(\beta'_{10}+\alpha'_{01})(\pd\pi \pd \xi)+\beta'_{01}(\pd \xi)^2\right]+\text{total derivative}+\ldots\,. 
\eea
The background has a ghost-like excitations  if
\be
K=\left[ \begin {array}{cc} 2\alpha'_{10}&\alpha'_{01}+\beta'_{10}\\\noalign{\medskip}
\beta'_{10}+\alpha'_{01} &2\beta'_{01}\end {array} \right]
\ee
has a {\it negative} eigenvalue. However, using Eqs.~\ref{cocd1}, \ref{cocd2} and the linear map Eq.~\ref{map}, we see that
\be
K=\left[ \begin {array}{cc} a'_{10}&a'_{01}\\\noalign{\medskip}
b'_{10}&b'_{01}\end {array} \right]=-\frac12 \sum_{1\leqslant i+ j\leqslant 4}\left(-\frac12\right)^{i+j}\frac{4!}{(4-i-j)!}\left[ \begin {array}{cc}i a_{ij}k_\pi^{i-1} k_\xi^j &ja_{ij}k_\pi^{i} k_\xi^{j-1} \\\noalign{\medskip}
ib_{ij}k_\pi^{i-1} k_\xi^j &jb_{ij}k_\pi^{i} k_\xi^{j-1}\end {array} \right]\,.
\ee
Differentiating  Eqs.~\ref{bg1} and \ref{bg2} then using Eq.~\ref{E=pdL}, one is able to show by direct comparison that
\be
K=-\frac12 \left[ \begin {array}{cc} \frac{\pd E_\pi}{\pd k_\pi} &\frac{\pd E_\pi}{\pd k_\xi} \\\noalign{\medskip}
\frac{\pd E_\xi}{\pd k_\pi} &\frac{\pd E_\xi}{\pd k_\xi} \end {array} \right]=\frac12 Hess(L)\,.
\ee
Recall that the  background solution contains a ghost if $K$ has a {\it negative} eigenvalue, which is equivalent to $Hess(L)$ also having a {\it negative} eigenvalue. Thus we have proven our claim\footnote{The analogous result holds for N galileons.}.

We can understand this result geometrically. The function $L(k_\pi, k_\xi)$ is a smooth polynomial function of two variables, whose extrema correspond to maximally symmetric vacua in our theory. Unstable vacua happen to be those for which the Hessian of $L$ has a negative eigenvalue on the solution. This just means that the corresponding extrema are saddles and maxima. Or put another way, stable vacua are simply {\it minima} of the action polynomial, $L$. Note that a vanishing eigenvalue will generically lead to strong coupling at all scales as the corresponding scalar degrees of freedom will have a vanishing kinetic term.

This geometrical insight enables us to make some generic statements about the spectrum of vacua in each case, using Morse theory. For example, if our galileon theory contains more than one  vacuum solution, they cannot all be stable. This is because Morse theory tells us that  you cannot have all extrema being minima if there is more than one extremum. This logic explains why the self-accelerating branch of the  DGP model must have a ghost-like instability~\cite{saghosts} if the normal branch is stable. 

The action polynomial is an extremely useful construct. One can use it to build interesting theories from the bottom up, ensuring stability conditions from the outset. This will be the approach we take in our detailed study of phenomenology \cite{otherpaper}. Given a suitable action polynomial, we can reconstruct the corresponding action for the relevant theory by computing the following coefficients:
\bea
a_{m,n}=(m+1)(\alpha_{m, n}+\beta_{m+1,n-1}) &=&-(-2)^{m+n}\frac{(4-m-n)!}{4!m!n!}~\frac{\pd^{m+n+1} L}{\pd k_\pi^{m+1} \pd k_{\xi}^{n}}\Bigg|_{k_\pi=k_\xi=0}+4\sigma \delta^{m}_0\delta_0^n \label{mixed3}\\
b_{m,n}=(n+1)(\beta_{m,n}+\alpha_{m-1,n+1})=  &=&-(-2)^{m+n}\frac{(4-m-n)!}{4!m!n!}~\frac{\pd^{m+n+1} L}{\pd k_\pi^{m} \pd k_{\xi}^{n+1}}\Bigg|_{k_\pi=k_\xi=0} \label{mixed4}
 \eea
where $m, n =0, 1, 2, \ldots$, and we recall that we define $\alpha_{-1, n}=\beta_{m, -1}=0$. Note that since $\pi {\cal E}_{m-1, n}-\xi {\cal E}_{m, n-1}$ is a total derivative for $n, m \geq 1$, we are free to set, say,   $\beta_{1,n}=\beta_{2,n}=\beta_{3,n}=\ldots=0$, without loss of generality. However, it is clear from Eq. \ref{mixed3} that this choice  subsequently fixes $\alpha_{m, n}$ uniquely.

The action polynomial can also be used to quickly calculate the equations of motion for fluctuations on a solution, given by Eqs.  \ref{eompids} and \ref{eomxids}. Given the action polynomial, we can compute the coefficients in the perturbation equations directly, using the following formulae, valid for $1 \leqslant m+n \leqslant 4$
\bea
a'_{m, n} &=&-(-2)^{m+n}\frac{(4-m-n)!}{4!m! n!}~\frac{\pd^{m+n+1} L}{\pd k_\pi^{m+1} \pd k_{\xi}^{n}}\\
b'_{m, n} &=&-(-2)^{m+n}\frac{(4-m-n)!}{4!m! n!}~\frac{\pd^{m+n+1} L}{\pd k_\pi^{m} \pd k_{\xi}^{n+1}}
 \eea
As it is much easier to work at the level of the action polynomial, these formulae  will prove invaluable to anyone wishing to study the phenomenology of maximally symmetric solutions and their fluctuations.

\section{Discussion} \label{sec:discuss}
We are now at the half way stage in our study of multiple galileon theory.  We will continue with a detailed analysis of phenomenology in our companion paper~\cite{otherpaper}, but for now let us take stock of where we are. We have generalised the original galileon theory~\cite{gal} to include  any number of coupled galileon fields.  We have focused our attention on the case of two galileon fields, one of which is coupled to the trace of the energy-momentum tensor. We call this version of the theory,  {\it bi-galileon theory}. Although the two galileons are coupled, perhaps even at quadratic order, we neglect any direct coupling between them and the graviton. This amounts to neglecting the backreaction of the scalars to background geometry.

Through the coupling to the energy-momentum tensor, the galileon fields lead to a long range modification of the gravitational force. We have argued that co-dimension two braneworld models with an infra-red modification of gravity will typically admit a decoupling limit corresponding to a bi-galileon theory. Indeed, we have shown explicitly how the cascading cosmology model~\cite{casccos} does exactly that.  Of course, strictly speaking the cascading cosmology model is a $5D$ model with co-dimension one branes \cite{casccos}, although it is expected to capture many of the salient features of the full $6D$ cascading gravity model \cite{casc1, casc2} which does contain co-dimension two branes. Our interest in galileon modifications of gravity need not rely on higher dimensional models of gravity. We can, in principle, reconstruct our bi-galileon theory along the lines of~\cite{covgal}, and develop a fully covariant model of two scalar fields coupled to gravity and to matter in four dimensions. While the Galilean invariance will be broken by this procedure, we expect it to be restored in the limit $M_{pl} \to \infty$, so we can still learn plenty about the model simply by looking at its bi-galileon limit.

Our construction of the most general bi-galileon theory is very similar to the single galileon case, with a straightforward extension to $N$ galileons. As was the case for the single galileon~\cite{gal}, there are only a finite number of terms you can write down in the equations of motion. This is because requiring Galilean symmetry, as well as no higher than second derivatives, forces us to build our equations of motion from scalar products of $\pd_\mu\pd_\nu \pi$ and $\pd_\mu\pd_\nu \xi$. These can be thought of as matrices from which we can only build a finite number of Cayley invariants. 

As a first step towards a study of the phenomenology, we have considered maximally symmetric vacua and their stability. We have found a very elegant way of describing the physics in terms of a particular function, the {\it action polynomial}, which is  closely related to the on-shell background Lagrangian. Stable vacua correspond to the minima of this polynomial. Computing its Hessian on a particular solution offers us a quick and easy stability test which we will make use of in the future~\cite{otherpaper}.  We can also apply Morse theory to understand the generic spread of stable and unstable vacua. For example, it becomes almost trivial to see that if there is more than one vacuum, they cannot all be stable.

Having motivated and constructed our general theory we conclude by referring the reader to our forthcoming companion paper~\cite{otherpaper} for a detailed discussion of phenomenology.  
\\~\\
{\it Note added:} Whilst this paper was in its very final stages of preparation,    \cite{pforms} appeared, in which the galileon model is generalised to mixed combinations of arbitrary p-forms. It is clear that the results of section \ref{sec:gentheory} of this paper can be derived from the formulae given in  \cite{pforms}, for the case of multiple scalars. Both this work and our own might be considered to be the natural generalisations of \cite{gengal}.

\subsection*{Acknowledgements}
We would like to thank Ed Copeland, Nemanja Kaloper, Christos Charmousis and Oriol Pujolas for useful discussions. AP is funded by a Royal Society University
Research Fellowship and SYZ by a CSRS studentship.

\appendix

\end{document}